\documentclass[article]{revtex4}
\usepackage{epsfig}
\usepackage{graphicx}%
\usepackage{dcolumn}
\usepackage{amsmath}

\makeatletter
\def\btt#1{\texttt{\@backslashchar#1}}%
\DeclareRobustCommand\bblash{\btt{\@backslashchar}}%
\makeatother

\begin{document}

\title{Features of Motion Around  Charged D-Stars}% Force line breaks with \\
\
\author{Xiang-hua Zhai}\email{kychz@shtu.edu.cn}

\author{Li-ping Fu}

\affiliation{Shanghai United Center for Astrophysics,Shanghai
Normal University, Shanghai 200234, China
}%

\date{\today}% It is always \today, today, but you may specify any date with \date.

\begin{abstract}
 The motion of light and a neutral test particle around the charged
D-star has been studied. The difference of the deficit angle of
light from the case in asymptotically flat spacetime is in a
factor $(1-\epsilon^2)$. The motion of a test particle is affected
by the deficit angle and the charge. Through the phase analysis,
we prove the existence of the periodic solution to the equation of
motion and the effect of the deficit angle and the charge to the
critical point and its type. We also give the conditions under
which the critical point is a stable center and an unstable saddle
point.
\end{abstract}

\pacs{11.10.Lm, 04.40.-b}% PACS, the Physics and Astronomy Classification Scheme.

\maketitle \textbf{PACS numbers}: 11.10.Lm, 04.40.-b

\vspace{0.4cm} \noindent\textbf{1. Introduction} \vspace{0.4cm}

Astronomical observations on the CMB anisotropy[1-3] and the
relation between red-shift and luminosity distance of SNeIa [4-6]
depicted that our universe is spatially flat, and about 23 percent
and 73 percent of energy density is resulted respectively from
cold dark matter and dark energy, respectively. The nature of
these substances are quite unusual and there is no justification
for assuming that it resembles known forms of matter or energy. On
the other hand, phase transitions of quantum fields in the early
Universe may produce various kinds of topological defects. The
idea that point-like defects known as monopoles ought to exist has
proved to be remarkably durable. The global monopole, which has
divergent mass in flat space is one of the most important defects.
The effects of gravity on the global monopole were firstly studied
by Barriola and Vilenkin [7]. When gravity is taken into account
the linearly divergent mass of a global monopole has an effect
analogous to that of a deficit solid angle plus that of a tiny
mass at the origin. It is shown that this small gravitational
potential is actually repulsive [8-12].

As possible candidates for the cold dark matter, various kinds of
cold stars such as Q-stars have been proposed [13-22]. Later, a
new class of cold stars named D-stars (effect stars) have been
proposed by Li et al. [23,24]. Compared to Q-stars , the D-stars
have a feature, that is, in the absence of the gravitational field
the theory has monopole solutions , which makes the D-stars behave
very differently from the Q-stars. Recently , these new objects
have farther been studied. Their possible interesting
astrophysical applications [25,26] and charged D-stars have also
been studied. In contrast to ordinary D-stars, there exists a U(1)
gauge field outside the charged D-stars [27].

In this paper, we will discuss the behavior of test particle and
geodesics in the exterior spacetime of the charged D-stars. With
the aid of phase-plane analysis, we carefully analyse the motion
of light and a test particle around a charged D-stars and show
some unique features of the behavior.

\vspace{0.4cm} \noindent\textbf{2. The charged D-Stars}
 \vspace{0.4cm}

To be specific, we shall work with a fixed model, where a global
$O(3)\times\ U_{em}(1)$ symmetry is broken down to $O(3)\times\
U_{em}(1)$. The Lagrangian density is (we work in units such that
$ \hbar=c=1$)[27],
\begin{equation}\label{lag}
L=g^{\mu\nu}D_\mu\phi
D_{\nu}\phi^*+\frac{1}{2}g^{\mu\nu}\partial_{\mu}
\sigma^a\partial_\nu\sigma^a-
  \eta^2\sigma^a\sigma^a\phi\phi^*-\eta
^2(\phi\phi^*)^2-\frac{\lambda^2}{8}(\sigma^a\sigma^a-\sigma_0^2)^2-\frac{1}{4}\
F^{\mu\nu}F_{\mu\nu}
\end{equation}

\noindent where $D_\mu=\partial_\mu+ieA_\mu$; $F_{\mu\nu}$ is the
electric-magnetic field vector, and $F_{\mu\nu}=\partial_\mu
A_\nu-\partial_\nu A_\mu$; $\sigma^a$ is a triplet of scalar
fields, isovector index $a=1,2,3$; $\phi$ is a complex scalar
field with the $U_{em}(1)$ symmetry $\phi\rightarrow
e^{-i\theta}\phi$. The field configuration describing a global
monopole is

\begin{equation}\label{config}
  \sigma^a=\sigma_0f(\rho)\frac{x^a}{\rho}, with    x^ax^a=\rho^2
\end{equation}

\noindent so that we will actually have a monopole solution if $f
\rightarrow 1$ at spatial infinity. There is the current
conservation ${j_{;\mu}}^\mu=0$, where

\begin{equation}\label{metric}
  j^\mu=i g^{\mu\nu}(\phi^*\frac{\partial\phi}{{\partial x}^{\nu}}-\frac{\partial\phi^*}{{\partial
  x^\nu}}\phi)
\end{equation}

This leads to the conservation of the electric charge

\begin{equation}\label{metric}
 Q=\int j^0\sqrt{-\det g} d^3x
\end{equation}

\noindent Introduce

\begin{equation}\label{scalareq}
\phi(\rho,t)=\frac{1}{\sqrt{2}}\sigma_0h(\rho)e^{i\sigma_0\omega
t}
\end{equation}

\noindent so that $f$ and $h$ are both dimensionless and real. As
the lowest energy solution of the theory, there should not exist
electric current in and around the charged D-stars, so there
should not exist the magnetic field. Therefore, the nonzero
component of $A_\mu$ is $A_0$, which can be chosen as

\begin{equation}\label{einsteineq}
A_0(\rho)=\frac{\sigma_0}{e}[\omega -g(\rho)]
\end{equation}

\noindent The general static metric with spherical symmetry can be
written as

\begin{equation}\label{metric}
  ds^2=B(\rho)dt^2-A(\rho)d \rho^2-\rho^2(d\theta^2+\sin^2\theta d\varphi^2)
\end{equation}

\noindent with the usual relation between the spherical
coordinates $\rho,\theta,\varphi$ and the "Cartesian" coordinates
$x^a$. Introducing a dimensionless $r\equiv   \sigma_0\rho$, from
the Lagrangian density (1) and the definitions for $f,h $ and $g$,
the scalar field equations can be obtained as

\begin{eqnarray}\label{scalareq}
 f^{''}+\Big[\frac{2}{r}+\frac{B^{'}}{2B}-\frac{A^{'}}{2A}\Big]f^{'}&=&Af\Big[\frac{2}{r^2}+\eta^2 h^2
  +\frac{1}{2}\lambda^2(f^2-1)\Big]\\\nonumber
   h^{''}+\Big[\frac{2}{r}-\frac{B^{'}}{2B}-\frac{A^{'}}{2A}\Big]h^{'}&=&Ah\Big[{\eta}^2f^2+ {\eta}^2 h^2
  -\frac{1}{B}g^2\Big]
\end{eqnarray}

\noindent the equation for electric-magnetic field is

\begin{equation}\label{scalareq}
   g^{''}+\bigg[\frac{2}{r}-\frac{B^{'}}{2B}-\frac{A^{'}}{2A}\bigg]g^{'}=A e^2 h^2 g
\end{equation}

\noindent where the prime denotes differentiation with respect to
$r$, and $A(r)$ and $B(r)$ are the metric fields. By expressing
the energy-momentum tensor of the system, one can reduce Einstein
equation: $G_{\mu\nu}=8\pi G T_{\mu\nu}$ .

The charged D-stars consists of three regions: an interior surface
and exterior, which can be discussed respectively [27]. For
considering the exterior region of the star, one formally solves
the Einstein equations of the static spherically symmetric metric
as follows [28]

\begin{eqnarray}\label{solution1}
 &&A^{-1}=1-\frac{8\pi G}{\rho}\int_0^\rho T_t^t\rho^2
 d\rho\\\nonumber
 &&B=\frac{1}{A(\rho)}\exp\bigg[8\pi G\int_\infty^\rho(T_t^t-T_\rho^\rho)A(\rho)\rho d\rho\bigg]
\end{eqnarray}

\noindent where the time coordinate has been rescaled so that
$B=A^{-1}$ as $\rho\rightarrow\infty$. In terms of the
dimensionless variable $r$ and another dimensionless quantity
$\epsilon$, the  Einstein equation can be formally integrated and
the solutions read as

\begin{equation}\label{solution3}
 A^{-1}(r)=1-\epsilon^2-\frac{2G\sigma_0 M_A(r)}{r}+\frac{G\sigma_0^2 Q_A^2(r)}{4\pi r^2}\nonumber
\end{equation}

\begin{equation}\label{solution4}
 B(r)=1-\epsilon^2-\frac{2G\sigma_0 M_B(r)}{r}+\frac{G\sigma_0^2 Q_B^2(r)}{4\pi r^2}
\end{equation}

\noindent where

 \begin{eqnarray}\label{ma}
&&M_A(r)=4\pi\sigma_0\exp[-\triangle(r)]\int_0^r
dr\exp[\triangle(r)]\times\nonumber\\&&\bigg\{f^2-1+r^2\bigg[\frac{1}{2B}g^2
h^2+\frac{1}{8}\lambda^2(f^2-1)^2+
    \eta^2(\frac{1}{2}f^2 h^2+\frac{1}{4}h^4)\bigg]+\frac{1}{2}r^2(1-\epsilon^2+\frac{G\sigma_0^2 Q_A^2}
    {4\pi r^2})(h^{'2}+f^{'2})\bigg\}\nonumber
\end{eqnarray}

\begin{eqnarray}\label{qa}
Q_A^2=16\pi^2r\exp[-\bigtriangleup_1(r)]\int_0^r r^2
dr\exp[\bigtriangleup_1(r)](1-\epsilon^2-\frac{2G\sigma_0M_A}{r})\frac{g^{'2}}{e^2
B}
\end{eqnarray}

\begin{eqnarray}\label{mb}
M_B(r)=M_A(r)\exp[\widetilde{\triangle}(r)]
+\frac{r(1-\epsilon^2)}{2G\sigma_0}\{{1-\exp[\widetilde{\triangle}(r)]}\}\nonumber
\end{eqnarray}
\noindent and
\begin{eqnarray}\label{qb}
Q_B^2=Q_A^2\exp[\widetilde{\triangle}(r)]\nonumber
\end{eqnarray}

\noindent in which

\begin{eqnarray}\label{delta}
&&\triangle(r)=\frac{\epsilon^2}{2}\int_0^r
dz(h^{'2}+f^{'2})z\nonumber\\
&&\triangle_1(r)=\frac{\epsilon^2}{2}\int_0^rdr\frac{g'}{e^2 B}r\\
&&\widetilde{\triangle}(r)=\epsilon^2\int_\infty^rdz(h^{'2}+f^{'2}+2A\frac{1}{2B}g^2
h^2)z\nonumber
\end{eqnarray}

\noindent If this convergence is fast enough in asymptotic
spacetime, $M_A(r)$and $M_B(r)$will also quickly converge to
finite values. Therefore, one can find the asymptotic expansions:

\begin{equation}\label{asymf}
  f(r)=1-\frac{1}{r^2}-\frac{3-2\epsilon^2}{2r^4}+O(r^{-6})\nonumber
\end{equation}

\begin{equation}\label{asyma}
  M_A(r)=M_A+\frac{4\pi\sigma_0}{r}+O(r^{-3})
\end{equation}

\begin{equation}\label{asymb}
  M_B(r)=M_A(r)(1-\frac{\epsilon^2}{r^4})+\frac{4\pi\sigma_0(1-\epsilon^2)}{r^3}+O(r^{-7})\nonumber
\end{equation}

Since the effective mass  $M_A(r)$ approaches very quickly its
asymptotic value $M_A$, it is a good approximation to take it as

\begin{equation}
M_A(r)=M_A+\frac{4\pi\sigma_0}{r}
\end{equation}
Therefore, to
investigate the motion of light and a test particle around a
charged D-star, we can take the metric coefficients as

\begin{equation}\label{solution5}
 A(\rho)^{-1}=B(\rho)=1-\epsilon^2-\frac{2G\widetilde{M}}{\rho}+\frac{2G\widetilde{Q}^2}
 {\rho^2}
\end{equation}

\noindent where, for convenience, we have rescaled
$\widetilde{M}=\sigma_0M_A$ and $\widetilde{Q}^2=\sigma_0^2
\frac{Q_A^2-32\pi^2}{8\pi}$.

\vspace{0.4cm} \noindent\textbf{3. The behavior of null geodesic
outside the charged D-stars :}
 \vspace{0.4cm}

The orbit of light outside of the charged D-star can be obtained by
solving the geodesic equation
\begin{equation}\label{geodesic}
  \frac{d^2x^{\rho}}{d\tau^2}+\Gamma^{\rho}_{\alpha\beta}\frac{dx^{\alpha}}
  {d\tau}\frac{dx^{\beta}}{d\tau}=0
\end{equation}

\noindent where $\tau$ is the affine parameter. But by using the
fact that $g_{ab}\frac{dx^a}{d\tau}\frac{dx^b}{d\tau}=0$ for null
geodesics and the constant of motion

\begin{equation}\label{constant1}
 E=B(r)\frac{dt}{d\tau}
\end{equation}

\noindent and

\begin{equation}\label{constant2}
  L=r^2\frac{d\varphi}{d\tau}
\end{equation}

\noindent one can, considering the motion confined on the
$\theta=\frac{\pi}{2}$ plane, easily obtain the equation for
geodesics as[29].

\begin{equation}\label{nullgeodesic}
\frac{1}{2}\dot{r}^2+\frac{1}{2}B(r)(\frac{L^2}{r^2})=\frac{1}{2}E^2
\end{equation}

\noindent Introducing the effective potential

\begin{equation}\label{effv}
 V_{eff}=\frac{1}{2}B(r)(\frac{L^2}{r^2})=\frac{L^2(1-\epsilon^2)}{2r^2}-
 \frac{L^2 G\widetilde{M}}{r^3}+\frac{GL^2\widetilde{Q}^2}{r^4}
\end{equation}

\noindent the geodesics of light becomes the same as that of a
neutral test particle with unit mass moving in the effective
potential [30]. Taking $G=1$ for convenience, it is easy to find
that the effective potential has a maximum

\begin{equation}\label{vmax}
 V_{max}=\frac{4L^2(1-\epsilon^2)^3\Big[3\widetilde{M}^2-4\widetilde{Q}^2(1-\epsilon^2)
 +\widetilde{M}\sqrt{9\widetilde{M}^2-16\widetilde{Q}^2(1-\epsilon^2)}\Big]}
 {\Big[3\widetilde{M}+\sqrt{9\widetilde{M}^2-16\widetilde{Q}^2(1-\epsilon^2)}\Big]^4}
\end{equation}

\noindent at

\begin{equation}\label{rcric}
r=r_c=\frac{3\widetilde{M}+\sqrt{9\widetilde{M}^2-16\widetilde{Q}^2(1-\epsilon^2)}}{2(1-\epsilon^2)}
\end{equation}

\noindent When $\frac{1}{2}E^2=V_{max}$, we will have
\begin{equation}\label{bcrit}
  b_{crit}=\frac{\Big[3\widetilde{M}+\sqrt{9\widetilde{M}^2-16\widetilde{Q}^2(1-\epsilon^2)}\Big]^2}
  {2\sqrt{2(1-\epsilon^2)^3\Big[3\widetilde{M}^2-4\widetilde{Q}^2(1-\epsilon^2)+\widetilde{M}
  \sqrt{9\widetilde{M}^2-16\widetilde{Q}^2(1-\epsilon^2)}\Big]}}
\end{equation}

\noindent where $b_{crit}$ is the critical value of $b$, which is
the generalization of the apparent impact parameter and is defined
as $b=\frac{L}{E}$. The capture cross section of the charged
D-star will be
\begin{equation}\label{crosssection}
\sigma_c=\pi
b_{crit}^2=\frac{\pi\Bigl[3\widetilde{M}+\sqrt{9\widetilde{M}^2-16\widetilde{Q}^2(1-\epsilon^2)}\Bigr]^4}
  {8(1-\epsilon^2)^3\Bigl[3\widetilde{M}^2-4\widetilde{Q}^2(1-\epsilon^2)+\widetilde{M}
  \sqrt{9\widetilde{M}^2-16\widetilde{Q}^2(1-\epsilon^2)}\Bigr]}
\end{equation}

It is not difficult to prove that the path of light has a turning
point at the largest radius, $R_0$, where
$\frac{dr}{d\varphi}|_{r=R_0}=0$. Using Eqs.(18)-(20), one can
obtain the equation which can give the relation between $R_0$ and
$b$
\begin{equation}\label{R0}
b^{-2}R_0^4-(1-\epsilon^2)R_0^2+2\widetilde{M}R_0-2\widetilde{Q}^2=0
\end{equation}

\noindent In order to compute the deflect angle of the light, we
rewrite the Eqs.(18)-(20) as
\begin{equation}\label{deflight}
 \frac{d\varphi}{dr}=[r^4b^{-2}-(1-\epsilon^2)r^2+2\widetilde{M}r-2\widetilde{Q}^2]^{-1/2}
\end{equation}
\noindent The change of light when passing a charged D-star should
be $\Delta\varphi=\varphi_\infty-\varphi_{-\infty}$. Considering
the symmetry, we have
\begin{equation}\label{angle}
\Delta\varphi=2\int_{R_0}^\infty\frac{dr}{[r^4b^{-2}-(1-\epsilon^2)r^2+2\widetilde{M}r-2\widetilde{Q}^2]^{1/2}}
\end{equation}

\noindent The deflection of light up to the first order of
$\widetilde{M}$ ($\widetilde{M}$ is small in the unit $G=1$) is
given by [29]:
\begin{equation}\label{dangle}
 \delta\varphi=\Delta\varphi-\pi\approx \widetilde{M}\frac{\partial(\Delta\varphi)}{\partial
 \widetilde{M}}\Bigg|_{\widetilde{M}=0}=\frac{4\widetilde{M}}{(1-\epsilon^2)^{3/2}R_0}
\end{equation}

\vspace{0.4cm} \noindent\textbf{4. The motion of timelike
particles around a charged D-star}
 \vspace{0.4cm}

From Eqs.(17)-(19), the equation of motion for a timelike test
particle can be expressed as
\begin{equation}\label{timelike}
 (\frac{L}{r^2}\frac{dr}{d\varphi})^2=\widetilde{E}^2-\widetilde{\mu}^2B(r)-\frac{L^2}{r^2}B(r)
\end{equation}

\noindent where $\widetilde{\mu}=\frac{\mu}{\sigma_0}$ and
$\widetilde{E}=\frac{E}{\sigma_0}$ are the rescaled mass and
energy of the test particle. Introducing $\chi=\frac{1}{r}$,
substituting it into Eq.(30) and then differentiating the equation
with respect to $\varphi$, one will obtain the following equation:

\begin{equation}\label{timelikenew}
\frac{d^2\chi}{d\varphi^2}=\frac{1}{p}-(1-\epsilon^2+\gamma)\chi+\alpha\chi^2+\beta\chi^3
\end{equation}

\noindent where $p$, $\gamma$, $\alpha$ and $\beta$  are
dimensionless parameters defined as:

\begin{eqnarray}
\frac{1}{p}&=&\frac{G\widetilde{M}\widetilde{\mu}^2}{L^2}\nonumber\\
\gamma&=&\frac{\widetilde{\mu}^2}{L^2}\frac{G\widetilde{Q}^2}{4\pi}
\\\alpha &=&
3G\widetilde{M}\nonumber\\
\beta &=& -2\frac{G\widetilde{Q}^2}{4\pi}\nonumber
\end{eqnarray}

Noting that Eq.(31) is a nonlinear differential equation, it can
be integrated formally as

\begin{equation}\label{inte}
\varphi-\varphi_0=\int_{\chi_0}^{\chi}\frac{d\chi}{\sqrt{\int_{\chi_0}^{\chi}2
[\frac{1}{p}-(1-\epsilon^2+\gamma)\chi+\alpha\chi^2+\beta\chi^3]d\chi+\dot{\chi}_0^2}}
\end{equation}

\noindent where $\dot{\chi}_0^2$ is the initial value of
$\frac{d\chi}{d\varphi}|_{\varphi=\varphi_0}$. However, it is
impossible to obtain the exact expression by integrating the above
equation. In the following, we will gain some qualitative property
of the system with the aid of phase-plane analysis without solving
the equation numerically. To do so, we introduce two parameters as
$x=\chi$ and $y=\frac{d\chi}{d\varphi}$ and the autonomous system
corresponding to Eq.(31) will be

\begin{eqnarray}\label{autos}
\frac{dx}{d\varphi} &=& f(x,y)=y\\\nonumber \frac{dy}{d\varphi}&=&
g(x,y)=\frac{1}{p}-(1-\epsilon^2+\gamma)\chi+\alpha\chi^2+\beta\chi^3
\end{eqnarray}

Now, we prove the existence of periodic solution. According to the
well known Bendixson's criterion[30], the equation of motion will
have periodic solution if the divergence of the functional vector
of the autonomous system is vanishing, i.e., $\nabla\cdot(f,
g)=0$. It is obvious that the functional vector $(f, g)$
corresponding to the Eqs.(34) satisfies the criterion and
therefore indicates that Eq.(31) has a periodic solution. One can
also find that the periodic solution exists when
$\epsilon=\gamma=\beta=0$ which is the case of an ordinary star in
asymptotically flat spacetime. This shows that the presence of
 constant and deficit angle and the electric charge will not exclude the existence of
periodic solution from the equation of motion.

Next, we analyze the critical points on the phase plane. The
critical point is $(x_0, 0)$, where $x_0$ satisfies
\begin{figure}
\epsfig{file=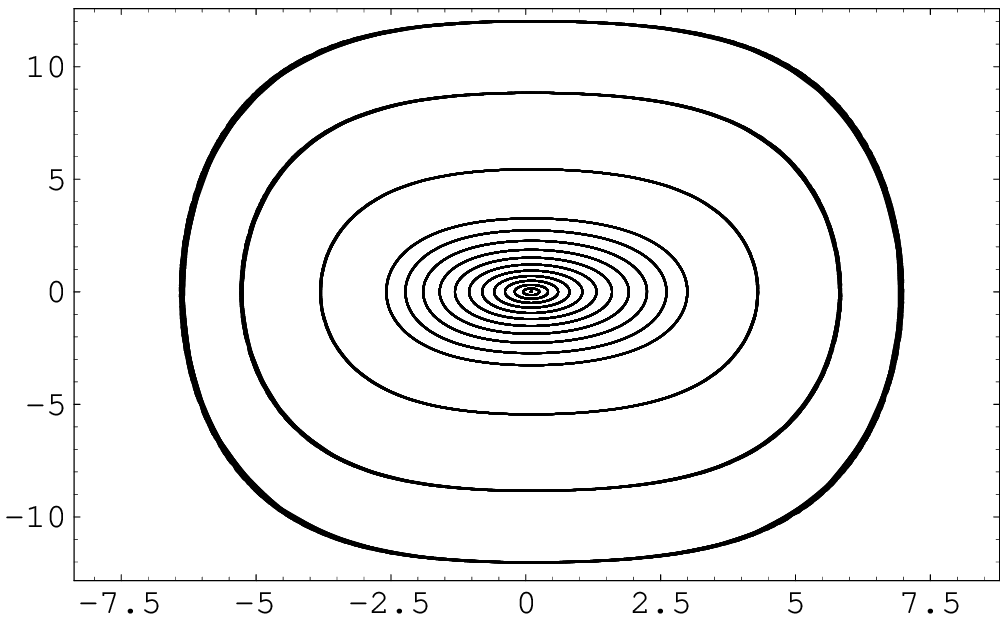,height=3in,width=4in} \caption{ The phase
graph when $\alpha=0.100$, $\gamma=0.100$, $\epsilon=0.010$,
$p=11$ and $\delta=0.001$, which corresponds to a charged case.}

\epsfig{file=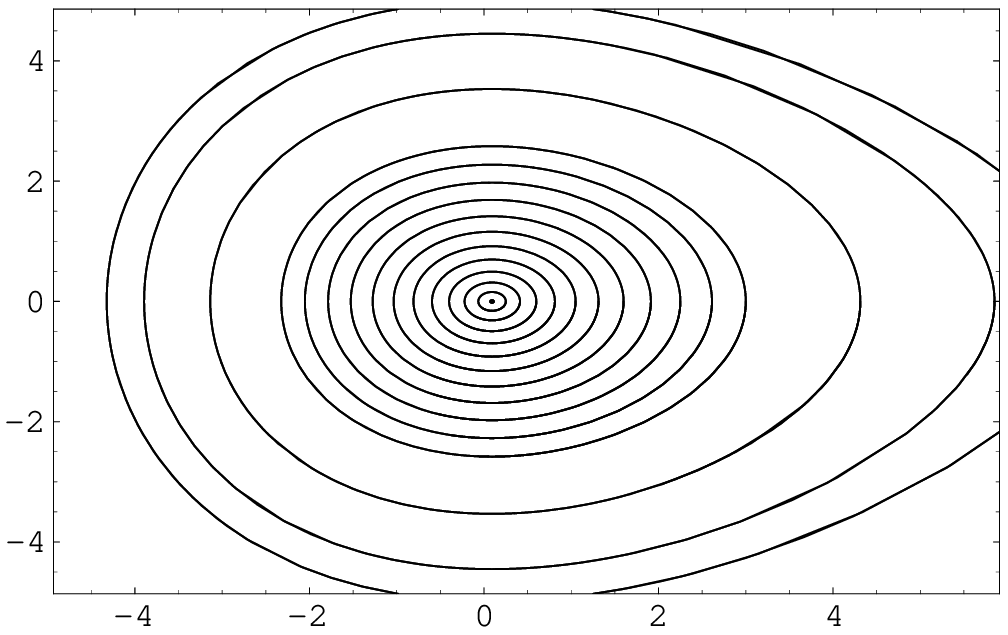,height=3in,width=4in} \caption{  The phase
graph when $\alpha=0.100$, $\gamma=0.000$, $\epsilon=0.000$,
$p=11$ and $\delta=0.000$, which corresponds to an uncharged
case.}
\end{figure}

\begin{equation}\label{critialp}
 \frac{1}{p}-(1-\epsilon^2+\gamma)x_0+\alpha x_0^2+\beta x_0^3=0
\end{equation}

\noindent To analyze the type of the critical point, we firstly
linearize the Eq.(34) and then do the translation $x=x-x_0$. Thus
the linearized equations should be:

\begin{eqnarray}\label{lautos}
\frac{dx}{d\varphi} &=& y\\\nonumber \frac{dy}{d\varphi}&=& \delta x
\end{eqnarray}

\noindent where
\begin{equation}\label{delta}
\delta=-(1-\epsilon^2+\gamma)+2\alpha x_0+3\beta x_0^2
\end{equation}

\noindent Using Eqs.(35) and (36), Eq.(37) could be rewritten as

\begin{equation}\label{delta1}
\delta=-\frac{3}{px_0}+2(1-\epsilon^2+\gamma)-\alpha x_0
\end{equation}

\noindent The eigenvalues corresponding to the system of equations will be

\begin{equation}\label{eigenv}
\lambda_{1, 2}=\pm\sqrt{\delta}
\end{equation}

\noindent The types of the critical point could be classified
according to the eigenvalues as following:

\noindent \textbf{I}. when $\delta>0$, we have
$\lambda_1<0<\lambda_2$, which indicates that the critical point
is an unstable saddle point. Considering the Eq.(38), this case
will correspond to the condition that

\noindent (1). $\alpha<0$ and $\Delta<0$.

\noindent (2). $\alpha>0$, $\Delta>0$ and
$\frac{(1-\epsilon^2+\gamma)-\sqrt{\Delta}}{\alpha}<x_0<\frac{(1-\epsilon^2+\gamma)+\sqrt{\Delta}}{\alpha}$.

\noindent (3). $\alpha<0$, $\Delta>0$ and
$x_0>\frac{(1-\epsilon^2+\gamma)-\sqrt{\Delta}}{\alpha}$,
\noindent where
$\Delta=(1-\epsilon^2+\gamma)^2-\frac{3\alpha}{p}$.

\noindent \textbf{II}. when $\delta<0$, we have two pure imaginary
eigenvalues $\lambda_{1, 2}=\pm\textbf{i}\sqrt{|\delta|}$, which
indicates that the critical point is stable center. Considering
Eq.(38), this case will correspond to the condition that

\noindent (1). $\alpha>0$ and $\Delta<0$.

\noindent (2). $\alpha>0$, $\Delta>0$ and
$x_0>\frac{(1-\epsilon^2+\gamma)+\sqrt{\Delta}}{\alpha}$,or$0<x_0<\frac{(1-\epsilon^2+\gamma)-\sqrt{\Delta}}{\alpha}$,

\noindent (3). $\alpha<0$, $\Delta>0$ and
$0<x_0<\frac{(1-\epsilon^2+\gamma)-\sqrt{\Delta}}{\alpha}$,

\noindent \textbf{III}. when $\delta=0$, we have $\lambda_{1,
2}=0$, which, together with the form of the autonomous system
Eqs.(36), indicates that the motion is uniformly on the lines
$y=Constants$ and all the the points on the lines $y=Constants$
are balanced positions. From Fig.1 to Fig.2, we show the phase
graph for different initial values and different parameters.

\vspace{0.4cm} \noindent\textbf{5. Discussion}
 \vspace{0.4cm}

We study the motion of light and a neutral test particle around a
charged D-star in this paper. We show that the deflect angle of
light is different from the case in asymptotically flat spacetime
for the existence of a factor of $(1-\epsilon^2)$ , which stands
for the feature of a D-star that the spacetime has a deficit
angle.

Using the phase-plane analysis, we investigate the qualitative
property of the dynamical equation controlling the motion of a
test particle around the charged D-star. We prove that the
equation of motion has a periodic solution, the position of whose
critical point and type on the phase plane can be affected by the
deficit angle and the charge. The conditions under which the
critical point is stable center and unstable saddle point
 have also been given .

\vspace{0.8cm} \noindent ACKNOWLEDGMENTS

This work was partially supported by the Foundation from Science
and Technology . Committee of Shanghai under Grant No.02QA14033
,and the Foundation from the Education Committee of Shanghai under
Grant No.01QN86.


\begin{thebibliography}{99}
\bibitem {NET}Netterfield C. B. et al. ,Astrophysics J.
\textbf{571}(2002)604.
\bibitem {B}Bennetl C. L. et al. , arXiv:astro-ph/0302207.
\bibitem {HAL}Halverson N. W. et al. , Astrophysics J.
\textbf{568}(2002)38.
\bibitem {RIE}Riess A. G. et al. , Astron. J.
\textbf{116}(1998)1009.
\bibitem {PER}Perlmutter S. et al. , Astrophysics J.
\textbf{517}(1999)565.
\bibitem {TO6}Tonry J. L. et al. , arXiv:astro-ph/0305008.
\bibitem {Barriola}Barriola M. and Vilenkin A. , Phys. Rev. Lett.
\textbf{63}(1989)341.
\bibitem {HARARI }Harari D. and Lousto C. , Phys.
Rev.\textbf{D42}(1990)2626.
\bibitem {SHI}Shi X. and Li X. Z. , Class. Quantum Grav.
\textbf{8}(1991)75.
\bibitem {L}Li X.Z. and Lu J. Z. ,Phys.
Rev.\textbf{D65}(2000)107501; Li X. Z. and Zhang J. Z. , Nuovo
Cim. \textbf{A105}(1992)1655; Li X. Z. Wang K. L. and Zhang J. Z.
, Nuovo Cim. \textbf{A84}(1984)311.
\bibitem {LXY}Li X.Z. and Hao J. G. ,Phys. Rev.
\textbf{D66}(2002)107501.
\bibitem {LIXY}Hao J. G. and Li X. Z. , Class. Quantum
Grav.\textbf{20}(2003)1703.
\bibitem {RUFFINI}Ruffini R. and Binazzola S. , Phys. Rev.
\textbf{187}(1969)1767.
\bibitem {COLPI}Colpi M. , Shapiro S.L. and Wasserman I. , Phys. Rev. Lett.,
\textbf{57}(1986)2485.
\bibitem {JETZER}Jetzer P. and Van Der Bij J. J. ,  Phys. Lett. B
\textbf{227}(1989)341.
\bibitem {LEE1}Lee T. D. , Phys. Rev. , \textbf{D35}(1987)3637.
\bibitem {FRIEDBERG1}Friedberg R. , Lee T.D. and PANG Y. ,  Phys. Rev.
\textbf{D35}(1987)3640.
\bibitem {FRIEDBERG2}Friedberg R. , Lee T.D. and PANG Y. ,  Phys. Rev.
\textbf{D35}(1987)3658.
\bibitem {LEE2}Lee T. D. and Pang Y. , Phys. Rev.
\textbf{D35}(1987)3678.
\bibitem {LYNN}Lynn B.W., Nucl. Phys.  \textbf{B321}(1989)465.
\bibitem {BAHCALL} Bahcall S. , Lynn B.W. and Selipsky S. B. , Nucl. Phys.
\textbf{B331}(1990)67.
\bibitem {COLEMAN}Coleman S. , Nucl. Phys. \textbf{B262}(1985)263.
\bibitem {LI1} Li X. Z. and Zhai X. H. , Phys. Lett.
\textbf{B364}(1995)212.
\bibitem {LI2}Li X. Z. , Zhai X.H. and Chen G. , Astropart. Phys.
\textbf{13}(2000)245.
\bibitem {LI3}Li J. M. and Li X. Z. , Chin. Phys. Lett.
\textbf{15}(1998)3.
\bibitem {LI4}Li X. Z. , Cheng H. B. and Kuo C. , Chin. Phys. Lett.
\textbf{18}(2001).
\bibitem {LI5}Li X. Z. , Liu D. J. and Hao J. G. , Sci.China,
\textbf{A45}(2002)520.
\bibitem{LAU}Landau L. D. , Lifshitz E. M. , 1971 The Classical
Theory of Fields, (Pergamon Press,  Oxford.1971).
\bibitem {WALD}Wald R. M. , General Relativity (Chicago , University of Chicago
Press),1984.
\bibitem{BENDIXSON}Sansone G. and Gonti R. , Non-Linear Differential
Equations (Oxford:Rergamon),1964.
\end{thebibliography}
\end{document}